%% file: main.tex
\documentclass[10pt,conference]{IEEEtran} % use the `journal` option for ITherm conference style
\IEEEoverridecommandlockouts
\usepackage{amsmath,amssymb,amsfonts}
\usepackage{algorithmic}
\usepackage{graphicx}
\usepackage[english]{babel}
\usepackage[colorinlistoftodos]{todonotes}
\usepackage{framed}
\usepackage{mdframed}
\usepackage{textcomp}
\usepackage{xcolor}
\usepackage[numbers]{natbib}
\usepackage{hyperref}
\usepackage[utf8]{inputenc}
\usepackage{rotating}
\usepackage{caption}
\usepackage{subcaption}
\usepackage{tikz}
\usepackage{multirow}
\newcommand{\numberedcircle}[1]{%
  \begin{tikzpicture}[baseline=(char.base)]
    \node[shape=circle, draw, inner sep=1pt] (char) {\small #1};
  \end{tikzpicture}}
\def\BibTeX{{\rm B\kern-.05em{\sc i\kern-.025em b}\kern-.08em
    T\kern-.1667em\lower.7ex\hbox{E}\kern-.125emX}}
\begin{document}

\title{Green Runner: A tool for efficient model selection from model repositories\\
% delete or comment-out the following line before submission
% {\footnotesize \textsuperscript{*}Note: Sub-titles are not captured in Xplore and should not be used}
}

\author{%%%% author names
    \IEEEauthorblockN{Jai Kannan\IEEEauthorrefmark{1}, Scott Barnett\IEEEauthorrefmark{1}, Anj Simmons\IEEEauthorrefmark{1}, Taylan Selvi\IEEEauthorrefmark{1}, Luís Cruz\IEEEauthorrefmark{2}}
    % duplicate the line above as many times as needed to list all authors
    %\\%%%% author affiliations
    \IEEEauthorblockA{\IEEEauthorrefmark{1}Applied Artificial Intelligence Institute, Deakin University, Geelong, Australia\\
Email: \{jai.kannan, scott.barnett, a.simmons, taylan.selvi\}@deakin.edu.au}%\\% first affiliation
    \IEEEauthorblockA{\IEEEauthorrefmark{2}Delft University of Technology, Delft, Netherlands\\
Email: l.cruz@tudelft.nl}%\\% delete this line if not needed
    % duplicate the line above as many times as needed to list all affiliations
    %%%% corresponding author contact details
    %\IEEEauthorblockA{email address or ORCID of corresponding author(s)}
}

\maketitle

\begin{abstract}
Deep learning models have become essential in software engineering, enabling intelligent features like image captioning and document generation. However, their popularity raises concerns about environmental impact and inefficient model selection. This paper introduces GreenRunnerGPT, a novel tool for efficiently selecting deep learning models based on specific use cases. It employs a large language model to suggest weights for quality indicators, optimizing resource utilization. The tool utilizes a multi-armed bandit framework to evaluate models against target datasets, considering tradeoffs. We demonstrate that GreenRunnerGPT is able to identify a model suited to a target use case without wasteful computations that would occur under a brute-force approach to model selection.
%GreenRunnerGPT provides an  comprehensive approach to reducing waste during model selection that considers both performance and environmental impact.
%The tool extends to cover tasks like fine-tuning and selection of machine learning API services.
%It offers an online resource for sustainable and efficient model selection.

% \textcolor{red}{A screencast of GreenRunnerGPT is available at youtube.com/... and an interactive online demo can be found at git@github.com:jai-kannan1184/GreenRunnerApp.git.}
\end{abstract}


\section{Introduction}

\label{sec:introduction}
\input{sections/introduction}

% \section{Motivation}
% \label{sec:motivation}
% \input{sections/motivation}

\section{GreenRunnerGPT}
\label{sec:Tool}
\input{sections/tool_description}

\section{Preliminary Evaluation}
\label{sec:eval}
\input{sections/preliminary_eval}

\section{Conclusion and Future Work}
\label{sec:conclusion}
\input{sections/conclusion.tex}

\bibliographystyle{IEEEtranN}
\bibliography{references}

\end{document}

%% file: sections/introduction.tex
Deep learning components have become an important aspect of software engineering. By using deep learning models (using a library or through an intelligent API web service), applications add intelligent features such as image captioning, image generation, and document generation. A growing concern with the popularity of deep learning is the environmental impact. The cost of operating a deep learning model is estimated to be  80\%--90\% of a model's total cost\footnote{https://www.forbes.com/sites/moorinsights/2019/05/09/google-cloud-doubles-down-on-nvidia-gpus-for-inference/} and inefficient use of computing resources contributes to the increased carbon footprint of machine learning (ML) \cite{strubell2019energy} research contributing to the global warming crisis. % \cite{masson2018global}
Selecting resource-efficient models in resource-efficient ways is paramount to reducing the environmental impact of deep learning. 

% training a single model has a carbon footprint equivalent to 5 family cars' lifetime emissions~\cite{strubell2019energy}. 

However, model selection is more than selecting the highest-performing model, tradeoffs need to be made dependent on the use case~\cite{Jabbarpour2021}. For example, a model designed to run on a drone to monitor crop health runs on resource-constrained hardware where a suboptimal model will consume higher compute draining the battery faster reducing flight-time. In comparison, a model operating in an autonomous vehicle has no hardware limitations although should have low latency to detect hazards in a timely manner, and a suboptimal model could be the difference between life and death \cite{inbook}.
%In the case of autonomous vehicles a 1-2\% increase in performance is the difference between life and death \cite{inbook}.

% These tradeoffs need to be considered during the model selection process. However, during model selection the data collection process is inefficient, i.e. collecting metrics on deep learning models requires compute resources.   

%Evaluation is a resource-intensive and time-consuming process as each model needs to be evaluated on each data point in the target dataset to determine how it will perform in operation.

\begin{figure}[tbp]
    \centering
    \includegraphics[height=2cm, width= \linewidth]{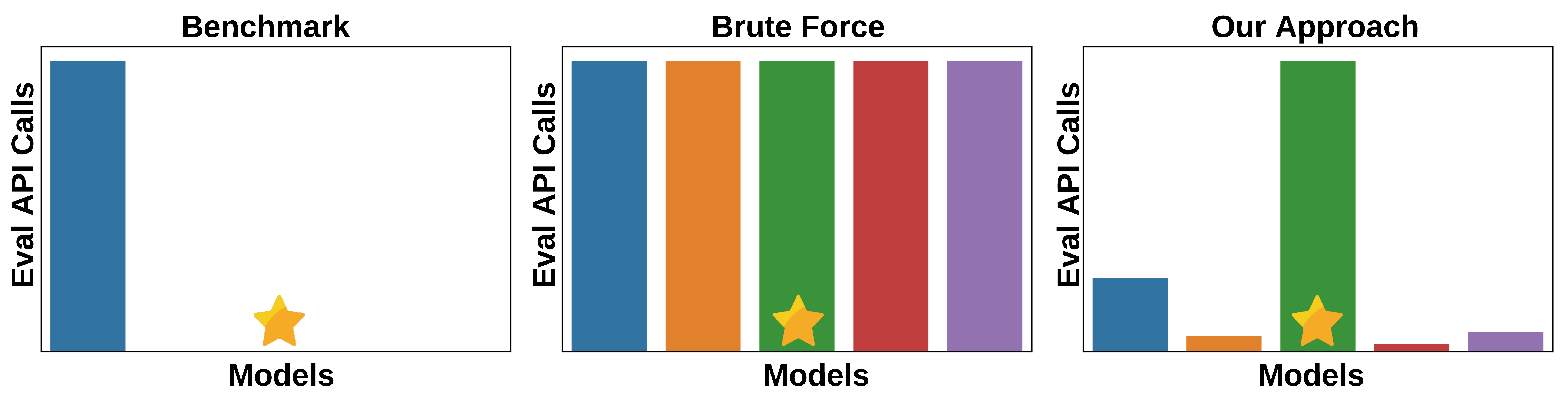}
    \caption{Model selection approaches: a) benchmark results (sub-optimal model selection), b) brute-force (wasteful API calls) and c) our proposed approach (balanced for target usecase)}
    
    % Comparative analysis of model selection process, with the ideal model indicated by the star. Ad-hoc approach (left) evaluates models using their performance on a benchmark (but this is not always representative of performance on the target data). Brute force approach (middle) evaluates using all data on all models (but this is wasteful). Our approach (right) efficiently identifies a model suitable for the target usecase.

    \label{fig:selection_methods}
\end{figure}

% moved to introduction section as a hack to get the tool figure to show at the top of page 2.
\begin{figure*}[h]
    \centering
    \includegraphics[width=\linewidth]{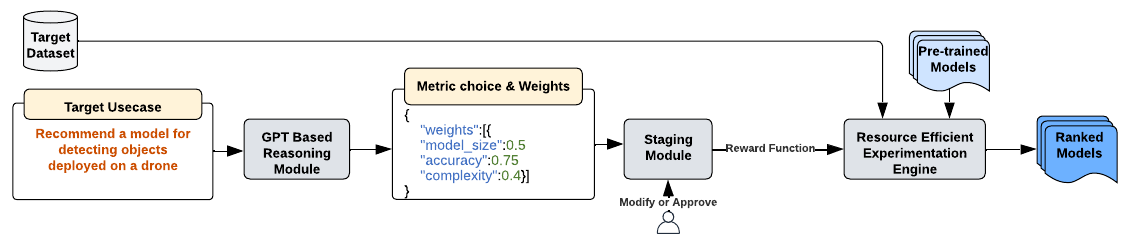}
    \caption{Overview of GreenRunnerGPT describing the internal processes and outputs from each process.}
    \label{fig:overview}
\end{figure*}

Model selection is currently an ad-hoc or wasteful process (see \autoref{fig:selection_methods}). The benchmark approach involves selecting models that perform well on a benchmark dataset and evaluating the best one on a target dataset. This approach finds suboptimal performing models as benchmark datasets are not representative of every target dataset \cite{barbu2019objectnet}. 
In contrast, the brute-force approach will result in a more accurate model but requires a full evaluation on all models which wastes resources (excessive number of model API calls).

We refer to this problem as the evaluation tax -- paid by either sacrificing model performance or wasting energy consumed evaluating deep learning models that are not part of the final application. The problem of evaluation tax is a growing concern due to the increasing number of available deep learning models. Hugging-face\footnote{https://huggingface.co/}, a popular model repository, has 198,613 models across 36 deep learning tasks all requiring evaluation and consideration of tradeoffs before use.

Prior work has proposed proxy metrics and selection strategies to reduce unnecessary computations during model selection, in particular, efficiently selecting models for the purpose of transfer learning or fine tuning on a target dataset \cite{Renggli2022, Li2021}. However, these approaches focus solely on identifying the highest-performing model(s) rather than considering quality tradeoffs during the model selection process.

%\cite{Renggli2022}
%\cite{kornblith2019better,Pandy2022,zamir2018taskonomy,ge2017borrowing}
%\cite{deshpande2021linearized,you2021logme}

%To bridge this gap, we created GreenRunnerGPT to create an optimised model selection process (see \autoref{fig:selection_ours}).

To bridge this gap, we propose a novel tool, GreenRunnerGPT, for efficiently selecting deep learning models \textit{balancing tradeoffs for a target usecase}.
%The core idea is to help developers select a suitable model for their dataset in an energy-efficient manner, taking into account the tradeoffs involved and the metrics that need to be optimized for the target use case.
To support this process, we use a large language model (LLM) to suggest weights to assign to each quality indicator (accuracy, model complexity, model size, etc.) given a plain-text description of the target use case. These are used as the reward function for a resource-efficient experimentation engine (built upon a multi-armed bandit framework) that efficiently evaluates models against the target dataset to determine the most promising models for the target use case.

%Large language models have code generation abilities and are trained on internet scale datasets with technical blogs, code repositories, and technical papers. Our hypothesis is that within a large language model is sufficient knowledge of quality tradeoffs for deep learning applications that can be used to generate suitable configuration. This reduces the effort to create evaluation configuration, and to quantify competing trade-offs. GreenRunnerGPT is implemented to support model selection with an architecture that supports fine-tuning, hyperparameter optimisation, and machine learning API web services. 

In this paper, we demonstrate how  GreenRunnerGPT helps to 1) select the right quality indicators for a particular scenario and 2) reduce the carbon footprint of selecting a model that excels in those indicators.

\begin{figure*}[h]
    \centering
    \includegraphics[width=\linewidth]{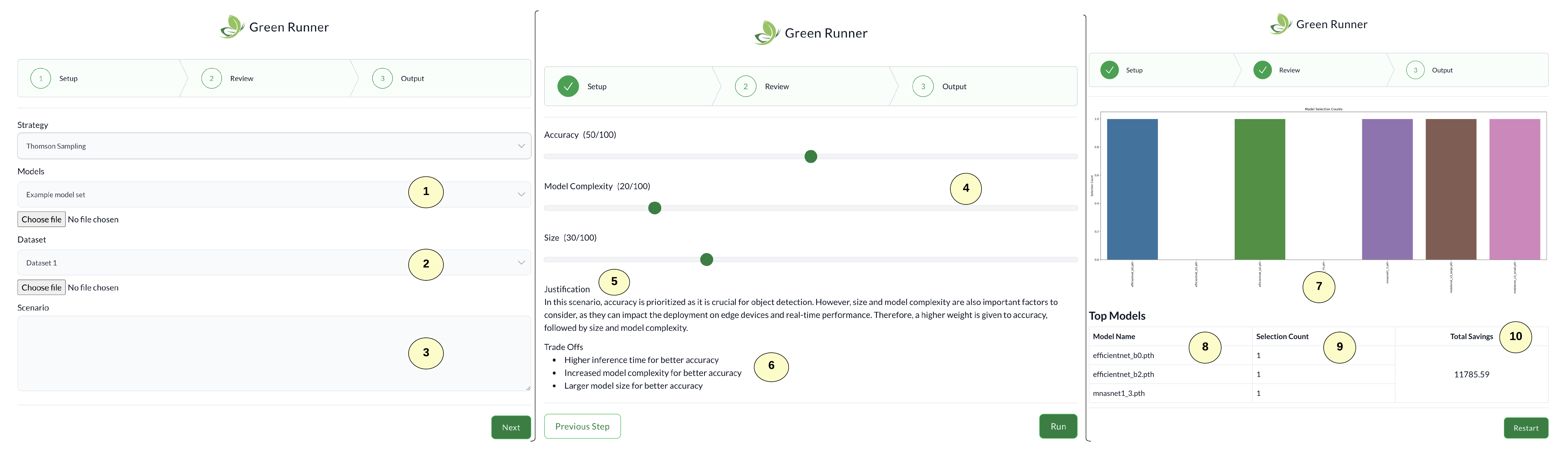}
    \caption{GreenRunnerGPT online tool: a) experiment setup page b) experiment staging page and c) Result analysis page}
    \label{fig:usage}
\end{figure*}

%% file: sections/tool_description.tex
% \subsection{Tool Overview}
In this section, we describe the core components of GreenRunnerGPT, shown in figure \autoref{fig:overview}. The core components are 1) a GPT Based Reasoning Module to suggest metric choice \& weights for a target use case, which form a reward function, and 2) a Resource Efficient Experimentation Engine that evaluates pre-trained models on a target dataset against using reward function to produce a set of top-ranked models for the target use case. 

\subsubsection{GPT Based Reasoning Module}

The integration of GreenRunnerGPT with an LLM enables users to provide a concise plain text prompt that describes the specific use case for the application which is the first step in identifying the model for a particular use case. For this paper, the LLM used by the reasoning module was GPT-4. Through the analysis of the provided description, the LLM generates a set of metrics and weights, which are used as part of a reward function for the model selection process. These weights are distributed across three fundamental metrics: i) Model accuracy, ii) Model size, and iii) Model complexity.
%Each weight is tailored to suit the unique requirements of the given application scenario.

Additionally, the LLM offers comprehensive justifications for the selected weights, optimizing them in accordance with the use case at hand. These metric weights serve as configuration parameters for resource-efficient evaluation algorithms, which effectively navigate the extensive model repository and select the most appropriate models. Prior to conducting the experiment, the staging module allows users to thoroughly examine and refine the metric weights, ensuring alignment with the intended use case.
%This iterative process enhances the overall efficiency of the model selection process, fostering optimal outcomes for the deployment of resource-efficient applications.

% The configuration module allows users to tailor the model analysis process to their specific needs. Users can select various constraints for the tool, including: i) the searching strategy for the bandits, ii) the maximum allowable model size on disk, iii) the average serving time for the models, iv) weightings for the classes in the dataset where the performance of a model on a particular class is prioritized during the search, and v) the average performance of the model.

\subsubsection{Resource Efficient Experimentation Engine}

GreenRunnerGPT incorporates a diverse range of multi-armed bandit (MAB) algorithms \cite{slivkins2019introduction} to optimize the utilization of the target dataset and minimize the number of model API calls made during the evaluation for efficient model evaluation. MABs, within the context of decision-making and optimization, involve a crucial tradeoff between exploration and exploitation. In this framework, an agent is confronted with a set of decision options or ``arms,'' each characterized by an unknown reward distribution. The agent's primary objective is to maximize cumulative reward over time by strategically selecting the most rewarding actions \cite{slivkins2019introduction}, this objective incentivizes the agent to identify the arm(s) with the highest reward as quickly as possible, and to discard evaluation of any arms that are found to be sub-optimal.

In the GreenRunnerGPT system, the models stored in the model repository are initialized as arms. To evaluate the top-performing arms within a given budget, GreenRunnerGPT employs a specialized reward function. This function leverages the metric weights generated during the previous step for the specific use case, calculating the reward each time an arm is pulled. The reward calculation incorporates the metrics and weights, including i) Model accuracy, ii) Model size, and iii) Model complexity,  defined in \autoref{eq:reward}.
\begin{equation}
\resizebox{\linewidth}{!}{%
\begin{tabular}{@{}c@{}}
$\text{reward} = \text{accuracy} \times \text{weight\_acc}$ \\[0.5em]
$- \left(\dfrac{\log(\text{size}) - \log(\text{min\_size})}{\log(\text{max\_size}) - \log(\text{min\_size})}\right) \times \text{weight\_size}$ \\[0.5em]
$- \left(\dfrac{\log(\text{complexity}) - \log(\text{min\_complexity})}{\log(\text{max\_complexity}) - \log(\text{min\_complexity})}\right) \times \text{weight\_complexity}$
\end{tabular}%
}
\label{eq:reward}
\end{equation}
% \begin{equation}
% \resizebox{\linewidth}{!}{%
% $\begin{aligned}
% \text{reward} &= \text{accuracy} \times \text{weight\_acc} \\
% &- \left(\dfrac{\log(\text{size}) - \log(\text{min\_size})}{\log(\text{max\_size}) - \log(\text{min\_size})}\right) \times \text{weight\_size} \\
% &- \left(\dfrac{\log(\text{complexity}) - \log(\text{min\_complexity})}{\log(\text{max\_complexity}) - \log(\text{min\_complexity})}\right) \times \text{weight\_complexity}
% \end{aligned}$%
% }
% \end{equation}

With the tradeoffs suggested by the LLM, the selection process varies based on the user-determined use case by taking into account the metric weights for that use case as part of the reward function used to optimise the selection of model(s). GreenRunnerGPT currently integrates three MAB selection strategies: i) Epsilon Greedy, ii) Upper Confidence Bound, and iii) Thompson Sampling. These algorithms effectively balance exploration (by evaluating new models) and exploitation (by selecting the model with the highest reward) according to the specific requirements of the use case.

Furthermore, the concept of a budget is introduced, representing the user-specified maximum number of calls to the evaluation API for a given experiment (a large budget will increase the chance of identifying the optimal model, while a small budget will reduce the number of evaluation API calls at the cost of sometimes selecting a sub-optimal model). As more data points are processed, the MAB algorithm continually updates its beliefs about the performance of each model using Bayesian inference. This iterative updating process accumulates evidence, allowing the algorithm to make informed decisions regarding which models are most likely to yield favorable results on the data. 

A ranked set of models identified through the experimentation are reported to the user, in particular the top 3 recommended models and their reported performance on the target dataset.

\section{Usage Example}

%To address Tom's challenge, we present an innovative tool designed to optimize model selection while minimizing resource consumption. 
Jane, a developer, works for a company specializing in machine learning-based systems for object detection on drones. The company has a repository of pre-trained models designed for this purpose. Jane's objective is to identify the most suitable model considering efficiency, complexity, and deployment constraints on the drone's hardware.

%\autoref{fig:set-up} shows the three main screens of GreenRunnerGPT.
%In this section, we also explain how Tom can use our tool to solve his problem in model selection.
% \begin{figure*}[htbp]
%     \centering
%     \includegraphics[width=\linewidth]{figures/Green-Runner-usage.png}
%     \caption{Screenshot of GreenRunnerGPT with the three screens for using the application.}
%     \label{fig:usageexample}
% \end{figure*}

For Jane to use our tool she is met with the setup screen \autoref{fig:usage}. In this screen, Jane needs to upload a set of models \numberedcircle{1} that need to be evaluated, choose the selection strategy and the target dataset \numberedcircle{2} that Jane has for her task. The next step is to provide a use case description in the prompt box \numberedcircle{3} in plain text. For example, \textit{Recommend a model for detecting objects deployed on a drone}. Once the data, model, and use case are described clicking run will send a query to the GPT-based reasoning module, which will output a set of metric weights along with the tradeoffs and justification for choosing the weights.

 The experiment staging screen displays the reasoning module outputs a set of metrics and weights that is recommended for the use case \numberedcircle{4}.
The reasoning module also provides a justification \numberedcircle{5} for choosing the metric weights which are based on  critical factors which influence the attributes of the selected models. These are features that are desired for performing the task described in the use case.
The trade-offs for the models are also described by the reasoning module in \numberedcircle{6}. This allows a developer like Jane to understand what are the aspects that she needs to consider while performing the experiment. 

At this stage, Jane also has an opportunity to review the weights set on the sliders in \numberedcircle{4}, and adjust them, as these weights influence the reward function of the experiment engine. Jane can then run the experiment and gather the results.
The experiment analysis screen \numberedcircle{7}  displays a graph of the selected models along with the top 3 models chosen \numberedcircle{8}, with their selection counts \numberedcircle{9} and the total computational savings in millions of multiply–accumulate operations (MMACs) \numberedcircle{10} as compared to brute force approaches.

%% file: sections/preliminary_eval.tex
\begin{table*}[]
\centering
\caption{Comparative analysis of selection methods for model selection.}
\begin{tabular}{|l|l|l|c|r|r|r|}
\hline
\multicolumn{1}{|c|}{Method} &
  \multicolumn{1}{c|}{Reward} &
  \multicolumn{1}{c|}{Most Selected Model} &
  \multicolumn{1}{l|}{\begin{tabular}[x]{@{}c@{}}Avg. Accuracy\\ on Target\end{tabular}} &
  Avg. Model Size &
  \begin{tabular}[x]{@{}c@{}}Avg. Model\\ Complexity\end{tabular} &
  \multicolumn{1}{l|}{Avg. Eval API Calls} \\ \hline
\multirow{2}{*}{Benchmark}    & Accuracy                     & maxvit\_t            & 0.29 & 124.5 MB & 19670 MMAC  & 100  \\ \cline{2-7} 
                              & Accuracy, Size, Complexity   & mobilenet\_v3\ & 0.17 & 22 MB    & 229 MMAC    & 100  \\ \hline
\multirow{2}{*}{Brute Force}  & Accuracy                     & regnet\_y\_128gf     & 0.45 & 2581 MB  & 127750 MMAC & 7455 \\ \cline{2-7} 
                              & Accuracy, Size, Complexity   & convex\_net\_small    & 0.29 & 114 MB   & 4470 MMAC    & 7455 \\ \hline
\multirow{2}{*}{Our Approach} & Accuracy                     & regnet\_y\_32gf     & 0.32 & 581 MB  & 32380 MMAC & 3360 \\ \cline{2-7} 
                              & Accuracy, Size, Complexity & swin\_v2\_s          & 0.30 & 199 MB   & 5790 MMAC    & 3045 \\ \hline
\end{tabular}
\label{table:Resultanalysis}
\end{table*}

We used the GPT Based Reasoning model (based on GPT-4) to suggest metric and weights for the use-case \textit{Recommend a model for detecting objects deployed on a drone}. As the response is non-deterministic, we executed the prompt 100 times resulting in average weights. The average weight suggested for each metric for this use case was: accuracy weight of 0.63, size weight of 0.25, and complexity weight of 0.21.

For the set of pre-trained models, we used 71 image classification models on PyTorch Hub\footnote{https://pytorch.org/vision/stable/models.html\#classification}, a widely-used open-source framework for machine learning applications. To allow for standardised comparison, we select models that are pre-trained and evaluated on the ImageNet dataset. Our target dataset was ObjectNet \cite{barbu2019objectnet}, specifically designed for evaluating vision models in real-world scenarios. It includes 113 classes that match those in ImageNet, from which we randomly sampled 100 images for evaluation.

\textit{Does GreenRunnerGPT find the most suitable model for inferred trade-offs?}
We conducted a comparative analysis of our approach with two other methods: the benchmark approach and the brute-force approach. Each experiment employed two combinations of metrics. The first combination focused solely on accuracy, while the second took into account accuracy, size, and model complexity. For each method we report the accuracy (on target), size (in MB) and complexity (MMACs per call) of the selected model. We also report the number of model evaluation API calls performed during model selection by each approach. As our approach may recommend different models each run, we perform 200 iterations and report the average \autoref{table:Resultanalysis}.
%in order to optimize the number of calls made to the evaluation API. %The results are presented in \autoref{table:Resultanalysis}.

The benchmark approach relies on the metrics provided in the model cards of the benchmark dataset (ImageNet) to select the model with the best performance, achieving accuracy scores of 0.80 \textit{on the benchmark dataset} when optimising for accuracy alone. However, the accuracy of the the selected model \textit{on the target dataset} was only 0.29. When considering accuracy, size, and complexity together, the selected model achieved an accuracy of 0.17 when evaluated on the target dataset, while also having a significantly smaller size. However, this approach is suboptimal as it fails to evaluate potentially superior models.

The brute-force approach involves passing all the data through each model under evaluation. Although this method allows us to identify the best models for each reward function, it is inefficient in terms of computational resources, resulting in a high number of API calls.

To evaluate our approach using the multi-armed bandit experiment engine for model selection, we chose Thomson Sampling as the selection strategy. When optimising for accuracy alone, our approach selects a model with an accuracy of 0.32 on the target, although it falls short compared to the brute-force approach.

When our approach is used with a reward function that considers accuracy, size, and complexity, the selected model demonstrates a higher accuracy of 0.30 compared to the brute-force and benchmark methods, albeit with a slightly larger size. However, the size remains within the parameters of the use case, while significantly reducing computational resources and the number of eval API calls, thereby expediting the selection process to identify a suitable model for the use case.

%% file: sections/conclusion.tex
% Deep learning growth leads to increased energy consumption and carbon emissions. Our paper proposes a green AI solution to the model selection problem using Bayesian bandits. Our tool balances exploration and exploitation and allows users to specify tradeoff constraints between metrics. It can promote environmentally-friendly ML and aid industries with limited resources. Customizable for different applications, it satisfies requirements while minimizing environmental impact.

% In conclusion, the rapid growth of deep learning has led to an increase in energy consumption and carbon emissions. Evaluating multiple machine learning models for a specific task can be an expensive process in terms of computational resources and time, leading to environmental concerns. In this paper, we proposed a solution to the model selection problem that incorporates tradeoffs and promotes green AI. Our technique uses Bayesian bandits to balance exploration and exploitation to select the best model using a limited number of data points. Our tool also provides an interface that allows users to specify constraints for tradeoffs between different metrics. We believe that this approach can promote the development of more environmentally-friendly machine learning applications and contribute to the growth of AI for industries with limited computational resources. By incorporating tradeoffs, our approach can also be customized for various applications, ensuring that the selected model satisfies different requirements while minimizing environmental impact.

GreenRunnerGPT proves to be an effective tool for efficiently selecting deep learning models for specific use cases. By leveraging suggested weights from a large language model and employing a multi-armed bandit framework, (near) optimal models are chosen while considering tradeoffs. Compared to benchmark and brute-force methods, GreenRunnerGPT outperforms both approaches in terms of optimality and efficiency respectively. The benchmark method selects suboptimal models as accuracy on source data cannot reliably predict accuracy on target data, while the brute-force approach identifies the optimal model, but wastes computational resources by evaluating all models. GreenRunnerGPT achieves high accuracy and acceptable model sizes by incorporating suggested weights. It significantly reduces computational resources, resulting in a faster and more efficient model selection process. This allows for informed decisions while minimizing resource consumption and the associated environmental impact of this, thus contributing to sustainability in deep learning research.

While the focus of this paper was on selecting a suitable pre-trained model from a model repository, as future work we intent to cover other tasks such as model fine-tuning and selection of machine learning API web services. These tasks are compatible with our existing architecture and reward function, but require different MAB strategies and interfaces respectively. As future work, we also plan to allow the GPT Based Reasoning Module to suggest additional metrics beyond accuracy, size and complexity. This is also compatible with the existing architecture, but requires modifications to the reward function and a method to handle cases where the reasoning module suggests an unimplemented metric is important for the use case (e.g. through automatically generating a code stub to measure the unimplemented metric and add it to the reward function).

%The experiments conducted on image classification models using ObjectNet as the target dataset demonstrate the superiority of GreenRunnerGPT in finding models with improved accuracy. Furthermore, its consideration of environmental impact contributes to sustainability in deep learning research.